\documentclass[aps,pra,twocolumn,groupedaddress,superscriptaddress,bibnotes,amsfonts,
citeautoscript,a4paper]{revtex4-1}
\usepackage{latexsym}
\usepackage{amsmath}
\usepackage{amssymb}
\usepackage{graphicx}
\usepackage[T1]{fontenc}
\usepackage[open]{bookmark}
\usepackage{hyperref}
\hypersetup{colorlinks=true,allcolors=blue}

\newcommand\ci{\mathrm{i}}
\newcommand\real{\mathrm{Re}}
\newcommand\imag{\mathrm{Im}}
\newcommand\pt{\mathcal{PT}}
\newcommand\omegax{\omega_{\mathrm{x}}}

\newcommand\omegaplus{\Omega_{+}}

\newcommand\omegamin{\Omega_{-}}

\newcommand\omegadif{\Omega_{\mathrm{R}}}
\newcommand\qrabi{Q_{\mathrm{R}}}
\newcommand\omegaexc{\omega_{\mathrm{x}}}

\newcommand\gammaexc{\gamma_{\mathrm{x}}}

\newcommand\omegacav{\omega_{\mathrm{c}}}

\newcommand\gammacav{\gamma_{\mathrm{c}}}

\newcommand\gammag{\gamma_{\mathrm{g}}}

\newcommand{\EP}{\mathrm{EP}^{*}}

\begin{document}

\title{Gain-compensated cavities for the dynamic control of light--matter interactions}

\author{Christos Tserkezis}%\,\orcidlink{0000-0002-2075-9036}}
\affiliation{POLIMA---Center for Polariton-driven Light--Matter Interactions, University of Southern Denmark, Campusvej 55, DK-5230 Odense M, Denmark}
\affiliation{Center for Nano Optics, University of Southern Denmark, Campusvej 55, DK-5230 Odense M, Denmark}
\author{Christian Wolff}%\,\orcidlink{0000-0002-5759-6779}}
\affiliation{POLIMA---Center for Polariton-driven Light--Matter Interactions, University of Southern Denmark, Campusvej 55, DK-5230 Odense M, Denmark}
\affiliation{Center for Nano Optics, University of Southern Denmark, Campusvej 55, DK-5230 Odense M, Denmark}
\author{Fedor A. Shuklin}%\,\orcidlink{0000-0003-0240-0035}}
\affiliation{POLIMA---Center for Polariton-driven Light--Matter Interactions, University of Southern Denmark, Campusvej 55, DK-5230 Odense M, Denmark}
\affiliation{Center for Nano Optics, University of Southern Denmark, Campusvej 55, DK-5230 Odense M, Denmark}
\author{Francesco Todisco}%\,\orcidlink{0000-0002-0188-6048}}
\affiliation{CNR NANOTEC, Institute of Nanotechnology, Via Monteroni, 73100 Lecce, Italy}
\affiliation{POLIMA---Center for Polariton-driven Light--Matter Interactions, University of Southern Denmark, Campusvej 55, DK-5230 Odense M, Denmark}
\affiliation{Center for Nano Optics, University of Southern Denmark, Campusvej 55, DK-5230 Odense M, Denmark}
\author{Mikkel H. Eriksen}%\,\orcidlink{0000-0002-0159-0896}}
\affiliation{POLIMA---Center for Polariton-driven Light--Matter Interactions, University of Southern Denmark, Campusvej 55, DK-5230 Odense M, Denmark}
\affiliation{Center for Nano Optics, University of Southern Denmark, Campusvej 55, DK-5230 Odense M, Denmark}
\author{P.~A.~D.~Gon\c{c}alves}%\,\orcidlink{0000-0001-8518-3886}}
\altaffiliation[Present address: ]{ICFO --- Institut de Ci{\`e}ncies Fot{\`o}niques, The Barcelona Institute of Science and Technology, 08860 Castelldefels (Barcelona), Spain}
\affiliation{Center for Nano Optics, University of Southern Denmark, Campusvej 55, DK-5230 Odense M, Denmark}
\author{N. Asger Mortensen}%\,\orcidlink{0000-0001-7936-6264}}
\affiliation{POLIMA---Center for Polariton-driven Light--Matter Interactions, University of Southern Denmark, Campusvej 55, DK-5230 Odense M, Denmark}
\affiliation{Center for Nano Optics, University of Southern Denmark, Campusvej 55, DK-5230 Odense M, Denmark}
\affiliation{Danish Institute for Advanced Study, University of Southern Denmark,
Campusvej 55, DK-5230 Odense M, Denmark}

\date{\today}

\begin{abstract}
We propose an efficient approach for actively controlling the Rabi
oscillations in nanophotonic emitter--cavity analogues based on the
presence of an element with optical gain. Inspired by recent
developments in parity-time ($\pt$)-symmetry photonics, we show that
nano- or microcavities where intrinsic losses are partially or fully
compensated by an externally controllable amount of gain offer unique
capabilities for manipulating the dynamics of extended (collective)
excitonic emitter systems. In particular, we discuss how one can
drastically modify the dynamics of the system, increase the overall
occupation numbers, enhance the longevity of the Rabi oscillations,
and even decelerate them to the point where their experimental
observation becomes less challenging. Furthermore, we show that
there is a specific gain value that leads to an exceptional point,
where both emitter and cavity occupation oscillate practically in
phase, with occupation numbers that can significantly exceed unity.
By revisiting a recently-introduced Rabi-visibility measure, we
provide robust guidelines for quantifying the coupling strength
and achieving strong-coupling with adaptable Rabi frequency via
loss compensation.
\end{abstract}

\maketitle

\section{Introduction}

The possibility to control the emission of light from natural
or artificial light sources at the nanoscale has been attracting
considerable interest~\cite{torma_rpp78,vasa_acsphot5,tserkezis_rpp83,
sun_nscale13,xiong_apl118,Dombi_RMP_2020,fernandez_acsphot5},
ever since Purcell showed that the dynamics of an emitter is strongly
affected by its environment~\cite{purcell_pr69}. The tremendous
opportunities that such a control enables have kept inspiring novel
designs for efficient cavities, appropriately tailored depending on
the emitters under consideration. Mirror cavities, the prototypical
templates in cavity quantum electrodynamics~\cite{walther_rpp69,
kockum_natrevp1}, have long been employed as the most straightforward
choice when considering atoms~\cite{haroche_phystoday42}, while Bragg
reflectors and photonic crystals constitute a natural option for
artificial emitters such as quantum wells or dots~\cite{weisbuch_prl69,
yoshie_nat432}. More recently, excitons in molecular aggregates or
transition-metal dichalcogenides (TMDs)~\cite{goncalves_aom8} have
been introduced as emitters with strong, collective (i.e., 
\emph{effective}) dipole moments, leading to the emergence, among
others, of plasmonic~\cite{bellessa_prl93,dintinger_prb71,zengin_prl114,
sugawara_prl97,chikkaraddy_nat535,santhosh_natcom7} and
Mie-resonant~\cite{tserkezis_prb98,todisco_nanoph9,castellanos_acsphot7,
shen_prb105} nanostructures as suitable effective cavities. What really
determines the appropriateness of the cavity in all these endeavours is
the linewidth of the emitter: the optical mode must be chosen to have a
comparable linewidth, and the coupling strength must exceed the damping
rates of the individual components~\cite{torma_rpp78}. Nevertheless,
in addition to this fundamental requirement, whatever other optical
properties might characterise the cavity can readily open new routes
for the manipulation of the strong-coupling
response~\cite{gurlek_acsphot5,stamatopoulou_prb102,stamatopoulou_nscale14,
vidal_sci373}.

A major boost in the quest for photonic templates with novel,
possibly ``exotic'' optical properties was recently provided by
the adoption of the concept of non-Hermiticity~\cite{bender_rpp70}.
While initially introduced in the context of nuclear physics,
non-Hermitian Hamiltonians eventually found a fertile playground
in photonics~\cite{elganainy_natphys14}, especially after the
realisation that they can still have real eigenvalues, as long as
they commute with the parity-time ($\pt$) operator~\cite{bender_prl80}.
Their appeal in photonics is due to the fact that $\pt$-symmetric
potentials can be achieved by incorporating a gain element ---widely
available in optics since the development of lasers--- that compensates
the intrinsic loss of the system~\cite{makris_prl100,guo_prl103}.
Explorations of $\pt$ symmetry have often revealed surprising
responses and intriguing novel designs, including unidirectional
propagation~\cite{ramezani_pra82,feng_natmat12,peng_natphys10,
huang_oex23} and cloaking~\cite{lin_prl106,sounas_prappl4},
lasers~\cite{feng_sci346,hodaei_sci346,gao_optica4},
gyroscopes~\cite{hokmabadi_nat576}, nanoantennas~\cite{sanders_nanoph9},
and potentially powerful sensors operating at the exceptional point
(EP)~\cite{wiersig_prl112,hodaei_nat548,chen_nat548,mortensen_optica,
Wiersig_pr}, i.e., the condition under which the eigenstates of the
Hamiltonian coalesce and the corresponding eigenvalues are equal.

Inspired by these developments and the richness of optical
phenomena that can benefit from loss--gain combinations, we
explore here the possibility of designing optical cavities
where the emergence and time evolution of strong coupling 
can be dynamically controlled through the externally provided
gain. We theoretically show that by increasing the amount of
gain it is possible to drive the exciton--cavity system so
as to increase the number of Rabi oscillations that can be
measured before damping prevails and the system completely
loses its coherence. We revisit a recently-introduced visibility
measure~\cite{todisco_nanoph9}, and provide a detailed 
gain--coupling map that quantifies the different coupling
regimes; based on this, one can manipulate the dynamics of
the system, and accelerate or decelerate the Rabi oscillations
so as to render their period more easily resolvable in experiments.
Backtracking the visibility map, one can then extrapolate to
infer about the properties of the coupled system in the absence
of gain. Finally, we show that by further increasing the gain along
specific paths, one can reach an EP, where the dynamics of the
system is completely altered, and the occupations of both 
emitter--cavity polariton modes significantly exceed unity and
oscillate nearly in-phase. This set of different behaviours
indicates that inclusion of gain can open new ways for tailoring
the dynamics of coupled emitter--cavity architectures, with both
fundamental understanding and practical applications in mind.
Analogous phenomena have already been observed for 
optical~\cite{ghosh_scirep6} and magnonic~\cite{sadovnikov_prappl18}
waveguides operating at the EP, but here we generalise the
treatment for any kind of collective polaritonic system.
We anticipate that related experiments can readily benefit
from the techniques developed in the context of $\pt$ 
symmetry~\cite{elganainy_natphys14}.

\section{Hamiltonian description}

\begin{figure}[h]
\includegraphics[width=\columnwidth]{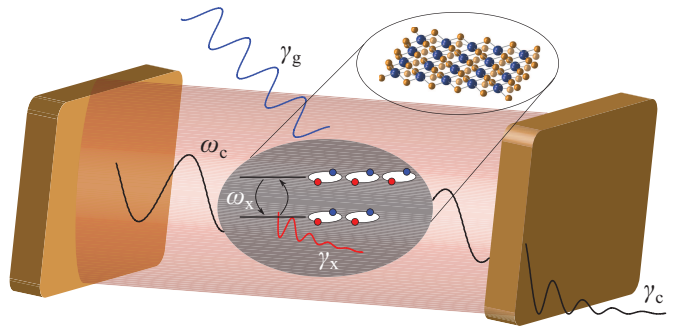}
\caption{
Schematic of the explored strong-coupling cavities: a typical cavity
formed by two mirrors, supporting a single optical mode with frequency
$\omegacav$ and damping rate $\gammacav$. Gain is provided to it
externally, at a rate $\gammag$. 
A QE system with a composite-bosonic character, formed by collective
excitonic resonances in extended systems such as TMDs (as shown in the
zoom-in sketch),
with transition frequency $\omegaexc$ and damping rate $\gammaexc$, is
placed between the mirrors and couples to the cavity mode with coupling
constant $g$.
}\label{fig1}
\end{figure}

We focus on extended, collective excitonic states
like those encountered in $J$-aggregates, individual organic
molecules, or TMDs, coupled to a (possibly open) micro- or
nano-cavity such as a distributed Bragg reflector, a metallic
nanoparticle, or simply a pair of mirrors, as suggested by the
schematic of Fig.~\ref{fig1}.
Expressing all the interactions in terms of the actual Hamiltonian
of the system becomes thus cumbersome, since one should include
interaction among excitons~\cite{agranovich_prb67}, together
with the appropriate Lindblad operators to account for both loss
and gain~\cite{franke_pra22}. We therefore restrict ourselves to
a toy-model description that essentially follows classical
coupled-mode theory~\cite{fan_josaa20}. We thus formulate the coupling problem in terms of a (semiclassical)
interaction Hamiltonian. To make the description as widely
applicable as possible, we consider a generic excitonic material,
modelled by a simple Lorentzian permittivity, with transition
frequency $\omegaexc$ and intrinsic linewidth $\gammaexc$,
coupled to a cavity with resonance frequency $\omegacav$ and
damping rate $\gammacav$. Within this  description, the dynamics
is then governed by the Schr\"{o}dinger-like equation
\begin{align}\label{Eq:Hamiltonian}
\begin{pmatrix}
\omegacav - \ci \dfrac{\gammacav}{2} + 
\ci \dfrac{\gammag}{2} & 
g \\
g & 
\omegaexc - \ci \dfrac{\gammaexc}{2}
\end{pmatrix} 
\begin{pmatrix}
\vphantom{\dfrac{x}{x}} a(t) \\ \vphantom{\dfrac{x}{x}} b(t)
\end{pmatrix} 
= \ci \frac{\partial}{\partial t} 
\begin{pmatrix}
\vphantom{\dfrac{x}{x}} a(t) \\ \vphantom{\dfrac{x}{x}} b(t)
\end{pmatrix}~,
\end{align}
where $g$ is the coupling constant and $\gammag$ is a possible
gain rate added to the cavity. What we aim to explore here is
if, and to what extent, the latter can be used as a means for
loss compensation that would eventually enhance the visibility
of Rabi-like oscillations in the strong-coupling regime. A
schematic of a typical cavity composed of two mirrors is shown
in Fig.~\ref{fig1}. A quantum emitter (QE), sketched in the figure
as a generic bosonic~\footnote{Excitons, being quasiparticles
made of two elementary fermions, are often well-described as
\emph{composite} bosons~\cite{CS2015} and thus exhibiting bosonic
character (bosonic commutation relations~\cite{Katsch_2018},
Bose--Einstein condensation~\cite{Pereira2021}, etc.) in the
low-density limit. In some circumstances, however, they can behave
more like fermions or even exhibit mixed bosonic and fermionic
effects~\cite{Combescot_2008,Katzer_arXiv23}.} system that could
correspond to the TMD shown in the zoom-in, characterised by an
``overall, effective emitter'' dipole moment,
is placed between the two resonators, and couples to a single cavity-mode
when the detuning $\omegaexc - \omegacav$ is sufficiently small. In addition
to the usual characteristics of such cavities, as encountered in quantum
optics, a gain element is also included here (Fig.~\ref{fig1}), e.g.,
by the inclusion of an active material that does not interact with the
QE, or via asymmetric pumping~\cite{li_acsphot9}. While, for our purposes,
it is sufficient to accept that some gain mechanism can exist, it
should be acknowledged that precise control of the gain rate is in
practice a challenging task, which requires carefully designed
experiments, appropriately adapted to the gain medium of choice
---which, ideally, should have a linewidth comparable to those of
the emitter and the cavity. Possibilities include electrochemical
doping for quantum dots~\cite{geuchies_nn15} or TMDs~\cite{morozov_aom9},
spatial modulation~\cite{keitel_nl21}, state-resolved optical
pumping~\cite{cooney_prl102} and host-guest chemistry~\cite{martini_natnano2}.

To describe the dynamics of the coupled system, we first assume for
simplicity a perfect frequency alignment between the cavity and the
exciton; this zero detuning is what most experiments try to achieve,
so as to better evaluate the coupling properties~\cite{wersall_nl17,
todisco_acsphot5,geisler_acsphot6}. Without loss of generality, we can
then measure all energies with respect to this frequency, i.e., set
$\omegacav = \omegax = 0$. We focus on time-harmonic solutions of the
form $e^{-\ci \omega t}$, and introduce frequencies and times normalised
to the linewidth of the exciton system: $\Omega \equiv 2\omega/ \gammaexc$
and $\tau \equiv \gammaexc t/2$. Likewise, we introduce the normalised
coupling $G \equiv 2g/\gammaexc$, and the normalised damping rate
$\Gamma \equiv (\gammag - \gammacav)/ \gammaexc$. In what follows,
we will explore the dynamics as $\Gamma$ varies from $-1$ (e.g., cavity
in the absence of gain and with a linewidth matching that of the QE)
to $+1$ (i.e., where the gain not only compensates the cavity losses,
but also exactly balances the broadening of the exciton), thus producing
long-lived Rabi-like oscillations. The time-harmonic solutions are now
governed by the dimensionless eigenvalue problem 
\begin{align}\label{Eq:eigenvalueproblem}
\begin{pmatrix}
\ci \Gamma & 
G \\
G &
-\ci
\end{pmatrix}
\begin{pmatrix}
a (\tau) \\
b (\tau)
\end{pmatrix}_{\pm}
= \Omega_{\pm} 
\begin{pmatrix}
a (\tau) \\
b (\tau)
\end{pmatrix}_{\pm}~,
\end{align}
whose diagonalisation yields the eigenfrequencies
\begin{align}\label{Eq:eigenfrequencies}
\Omega_{\pm} = 
-\frac{\ci}{2}
\left(1 - \Gamma \right) \pm 
\frac{1}{2} 
\sqrt{4G^{2} - \left(1 + \Gamma \right)^{2}}~.
\end{align}
In these dimensionless parameters, the Rabi-like frequency
$\omegadif = \omegaplus - \omegamin$ becomes
\begin{align}\label{Eq:omegadif}
\omegadif = 
\sqrt{4 G^{2} - \left(1 + \Gamma \right)^{2}}~,
\end{align}

\begin{figure*}[ht]
\includegraphics[width=2.0\columnwidth]{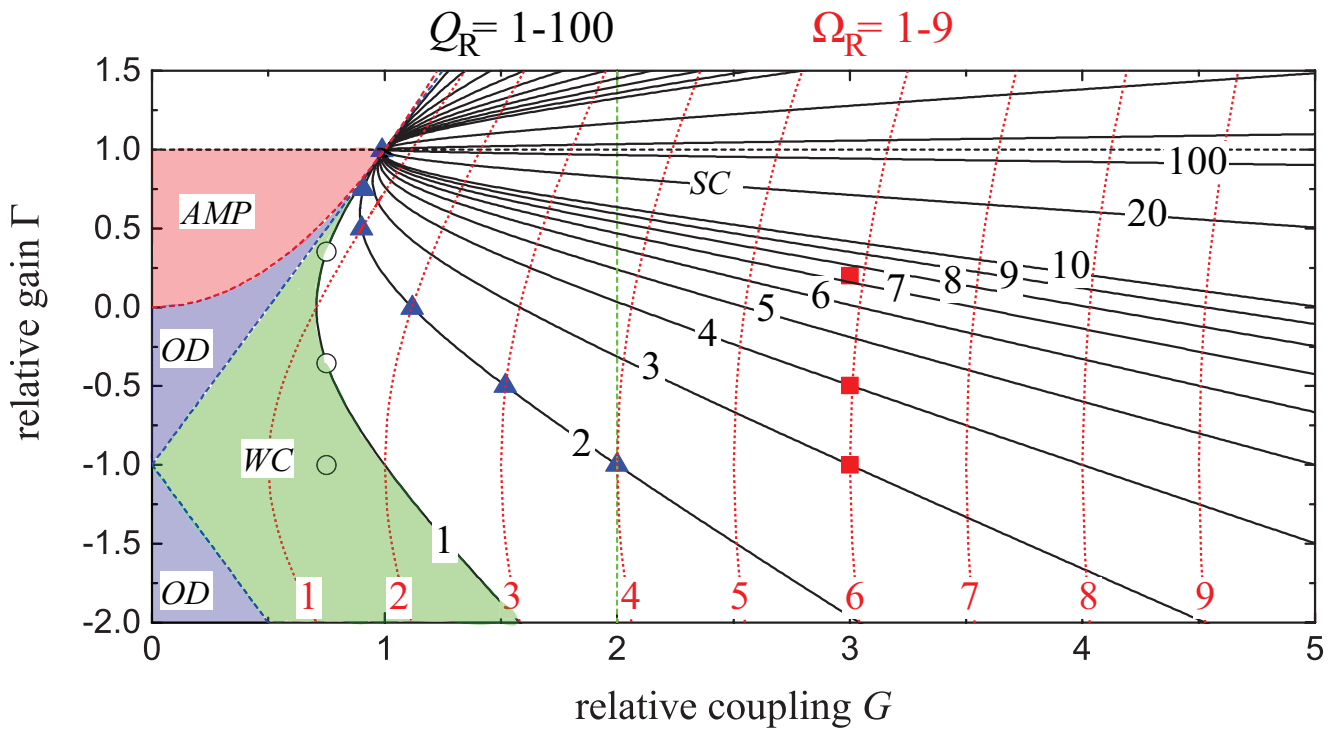}
\caption{Rabi-oscillation visibility quality factor ($\qrabi$) as a
function of the relative gain ($\Gamma$) and relative coupling ($G$).
Partial loss compensation occurs whenever $\Gamma > -1$, with
$\Gamma = 0$  corresponding to the case of a gain-balanced cavity
($\gammag = \gammacav$) and $\Gamma = 1$ to the case of fully
gain-compensated exciton--cavity system (i.e., $\gammag = \gammacav + 
\gammaexc$). When $2G = |\Gamma  + 1|$ (blue dashed lines), the two
solutions of Eq.~(\ref{Eq:eigenvalueproblem}) coalesce and give rise
to EPs; the one at $\Gamma = G = 1$ is particularly important and
all $\qrabi$ isocontours cross it. The blue-shaded region (OD) is
characterisedby critical damping, while the red-shaded region (AMP)
exhibits amplification. The weak-coupling regime (WC, green-shaded
region) and the strong-coupling regime (SC, white region) are
separated by the $\qrabi = 1$ curve. Black curves indicate
isocontours corresponding to different $\qrabi$, as indicated
by the labels in black font; red dotted curves
indicate isofrequency contours (constant $\omegadif$), with the
(normalised) values of the frequencies labelled in red font. Open
circles, blue triangles, and red squares correspond to specific $G$
and $\Gamma$ combinations discussed in the text.
}
\label{fig2}
\end{figure*}

\noindent which is a generalisation of the familiar result
$\omegadif = 2G$ (for $\Gamma = -1$ only). However, the introduction
of gain also opens the possibility for an EP, the condition being
$\Gamma = \pm 2G - 1$ (which corresponds to $\gammag = \pm 4g + 
\gammacav - \gammaexc$). This condition dictates the transition
between an oscillating and an overdamped (OD) system, shown by a
blue dashed line in the coupling map shown in Fig.~\ref{fig2}. When
this is fulfilled, the square root vanishes, leading to vanishing
splitting $\omegadif = 0$. In fact, the fulfillment of the
aforementioned condition, i.e., $2G = \big| \Gamma + 1 \big|$,
corresponds to a manifold of exceptional points, where the $\pt$
symmetry leads to coalescent eigenstates with entirely real eigenvalues
$\Omega_{\pm} = 0$. Hereafter, in our analysis we focus on one of such
EPs, namely that defined by $G = \Gamma = 1$, and henceforth denoted
$\EP$, which corresponds to a scenario where the gain exactly balances
the combined losses associated with the linewidths of the cavity and
emitter. As we shall see below, this point has intriguing consequences
for the dynamics of the ensuing light--matter interaction. Finally,
based on Eq.~(\ref{Eq:eigenfrequencies}), one can define the
amplification (AMP) region through $\imag \Omega_{\pm} > 0$.
One such AMP region, for which $\Gamma < 1$ and
$(1+\Gamma)^2 > 4G^2$, is highlighted with light-red colour in 
the top-left corner of
Fig.~\ref{fig2} ---it lies,
nevertheless, still in the OD regime.

\section{Visibility measure}

Before considering specific values of the normalised gain and
exploring how they affect the QE--cavity coupling, it is useful
to introduce a visibility measure for the Rabi oscillations in
terms of the quality factor ${\qrabi = \real \left[\omegamin - 
\omegaplus \right]/ \imag \left[\omegamin + \omegaplus \right]}$.
Such quality factors have already been introduced in recent literature
to deal with gainless strongly-coupled systems~\cite{yang_oex24,
todisco_nanoph9}, but here we generalise their applicability to the
case of cavities with gain. From Eq.~(\ref{Eq:eigenfrequencies}),
we straightforwardly obtain (while also assuming
$2G \geq \big|1+\Gamma\big|$)

\begin{table*}[t]
\begin{tabular}{|c|c|c|c|c|c|c|c|c|c|c|}
\hline
Ref.  & Cavity & Quantum Emitter & $g$ (meV) & $\gammaexc$ (meV) & $\gammacav$ (meV) & $G$ & $\Gamma$ & $Q_R$\\
\hline

\onlinecite{stuhrenberg_nl18} & Localised plasmon resonance & TMD exciton & 45 & 50 & 110 & 1.80 & -2.2 & 1.1 \\
\hline

\onlinecite{geisler_acsphot6} & Localised plasmon resonance & TMD exciton & 64 & 28 & 170 & 4.57 & -6.07 & 1.1 \\
\hline

\onlinecite{zengin_prl114} & Localized plasmon resonance & $J$-aggregate exciton & 81 & 100 & 109 & 1.62 & -1.09 & 1.6\\ 
\hline

\onlinecite{bouteyre_acsphot6} & Bragg mirror & Halide perovskite & 48 & 90 & 25 & 1.06 & -0.28 & 1.6 \\
\hline

\onlinecite{todisco_acsphot5} & Plasmon-lattice resonance & $J$-aggregate exciton & 137.5 & 80 & 200 & 3.44 & -2.50 & 1.9 \\
\hline

\onlinecite{lidzey_nat395} & Semiconductor microcavity & Organic semiconductor exciton & 80 & 90 & 22 & 1.77 & -0.24 & 2.8 \\
\hline

\onlinecite{chikkaraddy_nat535} & Localised-plasmon resonance & Dye-molecule exciton & 152.5 & 85 & 122 & 3.59 & -1.44 & 2.9 \\
\hline

\onlinecite{bellessa_prl93} & Surface-plasmon resonance & $J$-aggregate exciton & 90 & 50 & 70 & 3.6 & -1.4 & 3.0 \\
\hline

\onlinecite{wersall_nl17} & Localised plasmon resonance & $J$-aggregate exciton & 200 & 100 & 150 & 4 & -1.5 & 3.2\\ 
\hline

\onlinecite{dintinger_prb71} & Surface-plasmon resonance & $J$-aggregate exciton & ~125 & $\sim$ 0.66 & $\sim$ 140 & 380 & -213 & 3.4 \\ 
\hline 

\onlinecite{vasa_natphot7}  & Plasmon-lattice resonance & $J$-aggregate exciton & $\sim$ 350 & $\sim$ 20 & $\sim$ 200 & 37.5 & -10 & 4\\
\hline
\onlinecite{scalari_sci335} & THz metamaterial & 2DEG cyclotron transition & $\sim$ 4.1 & $\sim$ 0.41 & $\sim$ 0.12 & 20 & -0.29 & 31 \\
\hline
\onlinecite{dini_prl90} & Semiconductor microcavity & 2DEG intersubband transition & 7 & 5 & 15 & 5.6 & -3 & 45 \\
\hline
\onlinecite{rempe_prl64} & Superconducting microcavity & Atomic beam & 2.9$\times$10$^{-8}$ & 2.1$\times$10$^{-9}$ & 1.7$\times$10$^{-12}$ & 28 & -8.0$\times$10$^{-4}$ & 56 \\
\hline
\onlinecite{brune_prl76} & Superconducting microcavity  &  Circular Rydberg atoms & 1.0$\times$10$^{-7}$ & 2.1$\times$10$^{-11}$ & 3.0$\times$10$^{-9}$ & 1.0$\times$10$^4$  & -145 & 137 \\
\hline
\onlinecite{brennecke_nat450} & Mirror cavity & Bose--Einstein condensate & 1.3$\times$10$^{-3}$ & 1.2$\times$10$^{-5}$ & 5.4$\times$10$^{-6}$ & 213 & -0.4 & 298 \\
\hline
\end{tabular}
\caption{Table of strong-coupling experiments sorted by increasing $\qrabi$. The
similarity sign ($\sim$) is used when the data listed in the table are not mentioned explicitly by the authors of the corresponding reference, but roughly estimated in this paper. In all of these experiments, $\gammag = 0$, and thus $\Gamma = -\gammacav/\gammaexc$.}
\label{table1}
\end{table*}

\begin{align}\label{Eq:QR}
\qrabi &=
\frac{\sqrt{4 G^{2} - \left(1 + \Gamma \right)^{2}}}
{1-\Gamma} = 
\frac{\sqrt{\left(4g \right)^{2} - 
\left(\gammaexc + \gammag - \gammacav \right)^{2}}}
{\gammaexc - \gammag + \gammacav}~.
\end{align}
In the spirit of ring-down spectroscopy~\cite{maity_anchem93}, this
quality factor quantifies the number of ``round trips'', i.e., the
number of resolvable oscillations of light between the cavity and
the emitter. In passing, we emphasise how the linewidths are naturally
added up in accordance with Matthiessen's rule for the addition of
scattering rates~\cite{Ashcroft_Harcourt1976}. Introduction of this
measure for the visibility of Rabi-like oscillations allows us to rigorously
discuss the weak-coupling (WC) versus strong-coupling (SC) regimes.
Strong coupling occurs for $\qrabi > 1$, corresponding to $ G > 
\sqrt{\left(1 + \Gamma^{2} \right)/2}$ (white region in Fig.~\ref{fig2}),
which is perfectly in line with the more common definition that the
splitting should exceed the linewidth~\cite{torma_rpp78}. On the
other hand, for $\qrabi < 1$ the dynamics will have all the
characteristics associated with the WC regime
(light-green area in the left part of Fig.~\ref{fig2});
as $\qrabi$ approaches zero, the system enters either an overdamped
regime 
(OD, blue triangular regions at the leftmost end of
Fig.~\ref{fig2}), 
or the regime with net amplification, depending
on the relative gain. This is summarised in the parameter phase-space
of Fig.~\ref{fig2}, which provides a direct and intuitive guide for
manipulating the coupling via application of gain. The black curves
in the figure correspond to isocontours of $\qrabi$(values given with
numbers in black font), while the 
dotted red curves are isofrequency 
contours, for the $\omegadif$ values given
at the bottom of each curve in red font.

To evaluate the usefulness of $\qrabi$, it is insightful to examine
how state-of-the-art experiments from literature classify according
to this measure. Such a comparison is done in Table~\ref{table1} for
a variety of QEs and cavities, most of which employ surface plasmons
or localised plasmon resonances in nanoparticles as cavities, except
for a one-dimensional photonic crystal in Ref.~\cite{lidzey_nat395},
and quantum optical systems in Refs.~\cite{rempe_prl64,brune_prl76,
brennecke_nat450}. None of the experiments listed in the table have
used gain, meaning that the listed $\Gamma$ corresponds to the
normalised damping rate of the cavity alone (i.e., $\Gamma =
-\gammacav/ \gammaexc$). Using then this $\Gamma$ together with
$\gammaexc$, one can straightforwardly estimate if the usual
strong-coupling criterion
\begin{align}\label{Eq:SCcriterion}
\omegadif > \frac{\gammaexc + \gammacav}{2}
\end{align}
holds. This criterion is not normalised to any intrinsic property
of the system, and it is therefore difficult to use it to compare
different types of strong-coupling configurations, whereas $\qrabi$
is properly normalised and can then be applied to a wide variety of
systems. Hence, the predictions of Eq.~(\ref{Eq:SCcriterion}) do not
strictly follow the computed $\qrabi$ appearing in the table.

Inspecting Table~\ref{table1}, it appears that the highest $\qrabi 
(\geqslant 3)$ for nanophotonic systems are still achieved by
architectures employing $J$-aggregates coupled to plasmonic
systems~\cite{dintinger_prb71, wersall_nl17, bellessa_prl93}. 
This excellent performance is related to the high out-of-plane
dipole-moments of the $J$-aggregated molecules, and the intense
near fields provided by the plasmonic cavities. TMD-based systems,
on the other hand, seem to be the most poorly performing at the moment.
This has to do both with the fact that such activities have only
recently emerged ---leaving considerable room for improvement---
but also with the fact that the effective dipole moment associated
with excitons in these two-dimensional (2D) materials lie
predominantly in plane, making the coupling with any out-of-plane
cavity mode less efficient. In this respect, 2D halide perovskites,
with their dipole moments oriented out of plane~\cite{mao_jacs141},
might provide in the future an efficient alternative for 2D
polaritonics~\cite{bouteyre_acsphot6,su_natmat10}. Similarly,
high quality factors are also obtained for emitters based on 2D
electron gases (2DEGs), in the case of either intersubband
transitions~\cite{dini_prl90} or cyclotron transitions~\cite{scalari_sci335}.
Nevertheless, despite the recent advances in nanophotonics, the highest
performance in terms of $\qrabi$ lies still in the quantum-optical domain,
involving ultrahigh finesse (e.g. superconducting) cavities or Rydberg
atoms. This is to be expected, since in those platforms the linewidths of
both cavities and QEs can be very small, and all the experiments in
Refs.~\cite{rempe_prl64,brune_prl76,brennecke_nat450} have been carried
out at cryogenic temperatures. Of course, the requirement for such
conditions, and the costs that accompany them, was one of the main
motivation for shifting attention towards room-temperature nanophotonics
in the first place. One should thus always keep in mind the specific purpose
that any new strong-coupling configuration would serve, and find the best
balance between performance and cost. In passing, we should also mention
that the values of relative $\Gamma$ that we obtain by analysing the data
reported in Refs.~\cite{rempe_prl64,brune_prl76,brennecke_nat450} are
rather unconventional and unexpected (according to our previous
definition, values below $-1$ are expected for systems that are
not externally pumped, while values between $-1$ and $0$ correspond
to the presence of some kind of gain); these experiments typically
involve single atoms and/or photons, and adoption of our $\qrabi$
measure should be done more judiciously.

\section{Time evolution}

In order to further substantiate the usefulness of the above
visibility measure ---also in the time domain--- and to clearly
display that the parameter space of Fig.~\ref{fig2} indeed quantifies
the weak-versus-strong coupling regimes and the visibility of
Rabi oscillations in the latter case, we next consider the time
evolution of the cavity and emitter occupation numbers (herein
non-normalised due to the semiclassical incorporation of gain).
Starting with Eq.~(\ref{Eq:Hamiltonian}), the time evolution can be
solved straightforwardly (see Appendix~\ref{app:A}). For the initial
conditions of an empty cavity and an excited emitter, i.e.,
$a(0) = 0$ and $b(0) = 1$, we find
\begin{widetext}
\begin{subequations}\label{Eq:Dynamics}
\begin{align}
\left| a (\tau) \right|^{2} &= 
\frac{4 G^{2}}{\omegadif^{2}}
e^{- \frac{\omegadif \tau}{\qrabi}}
\sin^{2} (\tfrac{1}{2}\omegadif \tau) =
\frac{2 G^{2}}{\omegadif^{2}}
e^{- \frac{\omegadif \tau}{\qrabi}}
\left[1 - \cos (\omegadif \tau)\right]~,\\
%
%\left| b(\tau) \right|^{2} &= 
%e^{- \frac{\omegadif \tau}{\qrabi}}
%\left[ 
%\cos(\omegadif \tau) -
%\frac{\left(1  + \Gamma %\right)}{\omegadif}
%\sin(\omegadif \tau) 
%\right] + 
%\nonumber\\ & \quad  
%\left| a(\tau) \right|^{2}~.
\left| b(\tau) \right|^{2} &= 
e^{- \frac{\omegadif \tau}{\qrabi}}
\left[ 
\cos(\omegadif \tau) -
\frac{\left(1  + \Gamma \right)}{\omegadif}
\sin(\omegadif \tau) 
\right] + 
%\underbrace{\frac{4 G^{2}}{\omegadif^{2}}e^{- \frac{\omegadif \tau}{\qrabi}}\sin^{2} (\omegadif \tau/2)}_{\left|a(\tau) \right|^{2}}
\left|a(\tau) \right|^{2}~.
\end{align}
\end{subequations}
\end{widetext}
From these exact analytic expressions, it is unambiguously
clear that the occupation numbers oscillate with a period
governed solely by $\omegadif$, while only a number of
$\qrabi$ oscillations are visible, as a consequence of the
overall exponential decaying factor $e^{-\frac{\omegadif \tau}
{\qrabi}}$. Thus, our introduction of $\qrabi$ is more than a
convenient parameterisation ---it is the unique measure that 
emerges from a systematic dimensionless formulation of the problem.
We emphasise that, while initially $\left| a(0) \right|^{2} + 
\left| b(0) \right|^{2} = 1$, there is no such conservation after
a finite time. This is always anticipated in realistic systems
due to the dispersive and lossy nature of the cavity and the QE;
but in our case, because of the presence of gain, the sum of the
two occupation numbers can, and indeed does, exceed unity. In
what follows we will explore the dynamics of Eq.~(\ref{Eq:Dynamics})
for different limits of the parameter space depicted in Fig.~\ref{fig2},
and discuss how the presence of gain can drastically affect
the dynamics of the problem.

\subsubsection{Weak-coupling dynamics}

In the weak-coupling regime, with $\qrabi \ll 1$, the general
solution of Eq.~(\ref{Eq:Dynamics}) can be significantly simplified.
This is the Weisskopf--Wigner regime~\cite{Scully_Cambridge1997},
where the cavity exhibits an initial quadratic rise, 
\begin{subequations}
\begin{align}\label{Eq:DynamicsWeaka}
\left| a(\tau) \right|^{2} \approx 
\left(G \tau \right)^{2} 
e^{-\frac{\omegadif \tau}{\qrabi}}~,
\end{align}
where $G$ naturally determines the rate of growth, while the
accompanied initial dynamics of the QE is characterised by an
exponential decay, 
\begin{align}\label{Eq:DynamicsWeakb}
\left| b(\tau) \right|^{2} \approx
e^{- \frac{\omegadif \tau}{\qrabi}} = 
e^{-\left(1 - \Gamma \right) \tau}~.
\end{align}
\end{subequations}

\begin{figure}[h]
\includegraphics[width=1\columnwidth]{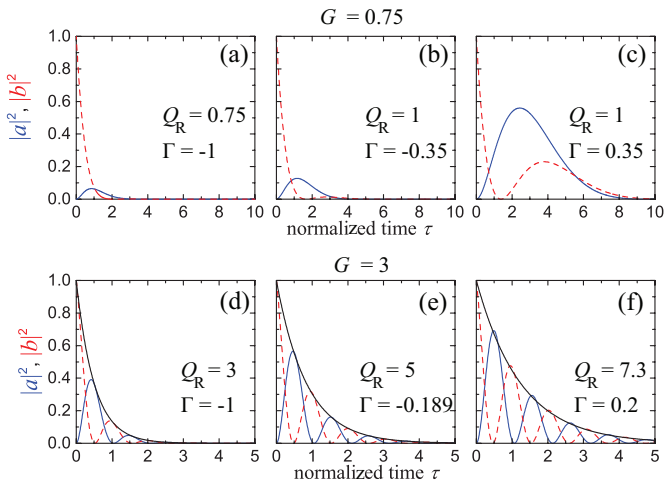}
\caption{
Upper panel: dynamics of the occupation numbers
$\left| b(\tau) \right|^2$ (red dashed curve) and 
$\left| a(\tau) \right|^2$ (blue solid curve), with
initial conditions $\left| b(0) \right|^2=1$ and
$\left| a(0) \right|^2=0$, for three cases with $G = 0.75$ 
(weak-coupling regime) and $\Gamma = -1.0$ (a), $\Gamma = -\sqrt{2}/4$
(b), and $\Gamma = \sqrt{2}/4$ (c) (see open circles in Fig.~\ref{fig2}).
Lower panel: similar dynamics for the strong-coupling regime
with $G = 3$ and $\Gamma = -1$ (d), $\Gamma = -0.189$ (e), and
$\Gamma = 0.2$ (f) (see red squares in Fig.~\ref{fig2}). Black
curves in the last three panels show the exponential decay
envelope of Eqs.~(\ref{Eq:DynamicsStrong}).
}\label{fig3}
\end{figure}

In Figs.~\ref{fig3}(a)-(c) we show this dynamics for three values of
$\Gamma$ along the $G = 0.75$ line in Fig.~\ref{fig2} (open circles).
In panel~(a), gain is completely absent $(\Gamma = -1)$, and the system
is entirely characterised by its intrinsic loss, leading to the anticipated 
exponential decay. Panel~(b) corresponds to $\Gamma = - \sqrt{2}/4\simeq
-0.35$ which, according to Eq.~(\ref{Eq:QR}) translates into the first of
the two points with $\qrabi = 1$, the one still dominated by loss (i.e.,
with $\Gamma < 0$). The emergence of a first oscillation in $\left| b(\tau) 
\right|^{2}$ is indeed (hardly) discernible in the red dashed
curve. Finally, panel~(c) corresponds to $\Gamma = \sqrt{2}/4\simeq  0.35$,
i.e. the second condition for which $\qrabi = 1$ in Fig.~\ref{fig2}; the
external gain has now started driving the system, so that a full oscillation
in $\left| b(\tau) \right|^{2}$ can be observed before its eventual decay.
According to Fig.~\ref{fig2}, it is feasible to drive the system into
the strong-coupling regime, just between these two points (before the
$\qrabi = 1$ curve backbends again for higher gain).  On the other hand,
keeping $G$ constant, say at $0.75$, and moving vertically in Fig.~\ref{fig2},
provides a recipe for tailoring the Purcell factor in the weak-coupling
regime. The Purcell factor expresses the acceleration of the spontaneous
emission rate in a cavity; the rate itself, is essentially proportional
to the square of the coupling strength, and inversely proportional to
the damping of the cavity~\cite{Agarwal_Cambridge20213}. Consequently,
for a fixed $G$, Fig.~\ref{fig2} suggests that spontaneous emission
rate increases, leading to a larger Purcell factor. Finally, it is
also important to observe that the period of the oscillations has
increased (in accordance with the shape of the isofrequency contours
in this region), and the two populations are more in phase; this will
become relevant again in the subsequent analysis.

\subsubsection{Strong-coupling dynamics}
Deep into the strong-coupling regime, where $G \gg 1$,
Eqs.~(\ref{Eq:Dynamics}) simplify to
\begin{subequations}\label{Eq:DynamicsStrong}
\begin{align}
\left| a(\tau) \right|^{2} & \simeq 
e^{-\frac{\omegadif \tau}{\qrabi}}
\sin^{2} \left( \tfrac{1}{2}\omegadif \tau \right) ~,\\
\left| b(\tau) \right|^{2} & \simeq 
e^{- \frac{\omegadif \tau}{\qrabi}}
\cos^{2} \left( \tfrac{1}{2}\omegadif \tau \right)~.
\end{align}
\end{subequations}
Here, we clearly see how the two occupation numbers evolve
fully out of phase, i.e., energy bounces back and forth
between the excitonic component and the cavity, while of course
still being exponentially damped over time. In turn, these
expressions imply that $\left| a(\tau) \right|^{2} + 
\left| b(\tau) \right|^{2} \simeq e^{- \frac{\omegadif \tau}{\qrabi}}$.
This is the common strong-coupling dynamics~\cite{Scully_Cambridge1997}.
Here, it is immediately evident that $\qrabi$ quantifies the number
of oscillations before reaching full relaxation: what the addition
of gain achieves is to increase the number of oscillations that
are visible before complete decay. Corresponding results are
shown in the lower half of Fig.~\ref{fig3}, panels~(d-f), for
$G = 3$ (red squares in Fig.~\ref{fig2}). When the system is not
compensated [$\Gamma =-1$, panel~(d)], three oscillations are observable
in the dynamics (the third only just). By increasing the amount of gain
provided to the system, one can increase the longevity of the Rabi
oscillations, and move vertically in the map of Fig.~\ref{fig2} (red
squares), leading to more oscillations (still dominated by the same
Rabi frequency, unlike in the previous case, as one can anticipate
by observing the corresponding isofrequency contour which is nearly
vertical) to be observed. The anticipated exponential decay is
demonstrated in all three cases by the black curves.

\subsubsection{Coalescing dynamics}

\begin{figure}[b]
\includegraphics[width=1\columnwidth]{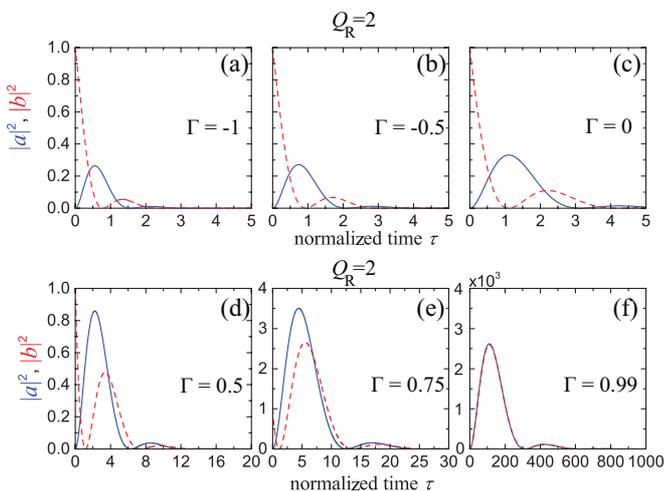}
\caption{
Dynamics of the occupation numbers $\left| b(\tau) \right|^{2}$
(red dashed curve) and $\left| a(\tau) \right|^{2}$ (blue 
solid curve), with initial
conditions $\left| b(0) \right|^{2} = 1$ and $\left| a(0) \right|^{2} = 0$,
for six cases on the $\qrabi = 2$ contour with $\Gamma = -1$ (a),
$\Gamma = -0.5$ (b), $\Gamma = 0$ (c), $\Gamma = 0.5$ (d),
$\Gamma = 0.75$ (e), and $\Gamma = 0.99$ (f), see blue triangles in
Fig.~\ref{fig2}.
}\label{fig4}
\end{figure}

A less explored special case, that only becomes relevant in the
kind of cavities with gain that we study here, concerns the dynamics
in the vicinity of an EP associated with $\pt$ symmetry. Looking at
Eqs.~(\ref{Eq:Dynamics}), we notice that, while  $\left| a(\tau) \right|^{2}
\propto \omegadif^{-2}$, the $\left| b(\tau) \right|^{2}$ part contains
terms scaling as $\omegadif^{0}$, $\omegadif^{-1}$, and $\omegadif^{-2}$.
Approaching $\EP$ (i.e., the EP for which $G = \Gamma = 1$), where
$\omegadif \to 0$, the $\omegadif^{-2}$ contribution dominates, and
consequently $\left| b(\tau) \right|^{2} \to \left| a(\tau) \right|^{2}$.
This implies an intriguing in-phase time evolution of the occupation numbers,
which is a remarkable consequence of $\pt$ symmetry and the coalescing
eigenstates at EPs. 

To see how the dynamics changes as the $\EP$ is approached, we
follow in Fig.~\ref{fig4} the $\qrabi = 2$ curve for
increasing $\Gamma$ (see blue triangles in Fig.~\ref{fig2}).
In the top three panels~(a-c), the composite cavity is still
lossy overall. In all cases, two oscillations are observed
(the second one barely) as expected, but their period increases
as the provided gain increases, in accordance with the behaviour
of the weakly-coupled system of Fig.~\ref{fig3}. This behaviour
could be immediately anticipated based on the isofrequency contours
of Fig.~\ref{fig2}: following the $\qrabi = 2$ line means crossing
several of the dotted red contours, with the frequency
decreasing as gain increases. In the lower panels~(d-f) the system
is practically externally driven by the gain. As the gain increases
and the $\EP$ point is approached, the Rabi oscillations are further
decelerated, both populations increase beyond the initial
condition of unity, and their phase difference becomes ever
smaller. Near $\EP$ (which cannot be reached exactly, so we
can only approach it adiabatically), both populations oscillate
practically in phase and coincide in their maximum values; the
system is driven externally, and has approached the net
amplification regime.

\section{Rabi-oscillation retrieval}

\begin{figure}[t]	
\includegraphics[width=1\columnwidth]{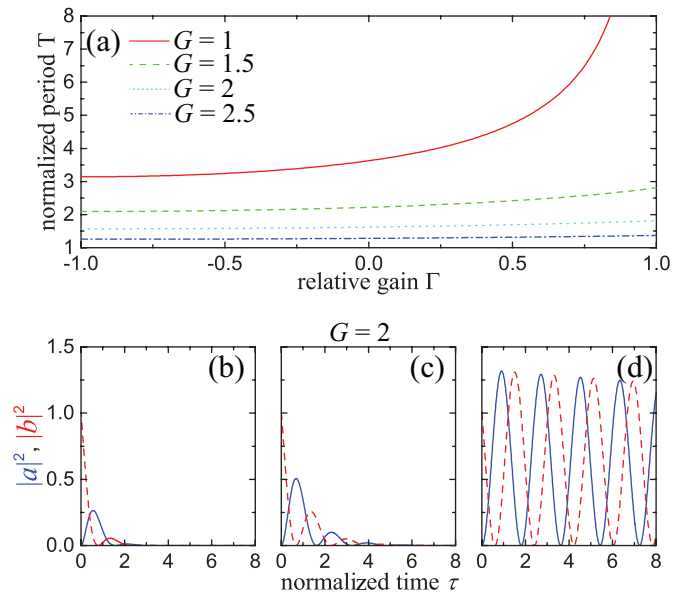}
\caption{
(a) Normalised period of the Rabi oscillations for a QE--cavity
system with $G$ equal to $1$ (solid red line) to $1.5$ (dashed
green line), $2$ (light-blue dotted line), and $2.5$ (dashed-dotted
blue line) as a function of the externally provided normalised gain
$\Gamma$. The period clearly follows an inverse square-root law,
in agreement with Eq.~(\ref{Eq:omegadif}).
(b)-(d) Time dynamics of the occupation numbers 
$\left| b(\tau) \right|^{2}$ (red dashed curves) and
$\left| a(\tau) \right|^{2}$ (blue solid curves), with initial
conditions $\left| b(0) \right|^{2} = 1$ and $\left| a(0) 
\right|^{2} = 0$, for $G = 2$ and $\Gamma = -1$ (b),
$\Gamma = 0$ (c), and $\Gamma = 0.99$ (d).
}\label{fig5}
\end{figure}

The gain-induced changes in the dynamics discussed above
suggest that one might be able to use gain to characterise
the time evolution of the system even if its initial Rabi
oscillations are too fast to be experimentally traceable [e.g., as
in typical plasmon-exciton coupling systems where $\omegadif$ can
be substantial (even though the corresponding $\qrabi$ are often
modest due to the sizable $\gammacav$)].
To this end, we plot in Fig.~\ref{fig5} the period of the
oscillations as a function of the externally provided gain,
for couplings ranging from $G = 1-2.5$. As the relative gain
increases, one moves vertically along Fig.~\ref{fig2}, meaning
that $\qrabi$ is expected to increase rapidly, especially for
relatively small $G$, for which the isofrequency contours are
more curved. At the same time, following the previous discussion,
the period of the oscillations is also expected to increase,
since it is just $T = 2\pi/\omegadif$, with $\omegadif$ given
by Eq.~(\ref{Eq:omegadif}). This is indeed shown in
Fig.~\ref{fig5}(a), for four cases of relatively small
(yet larger than unity) $G$. One immediately observes that
the weaker the coupling, the more it can be affected by the
exertion of gain, again in agreement with the isofrequency
contours of Fig.~\ref{fig2}. Three typical examples of the
dynamics are shown in panels~(b-d), for $G = 2$ (along the
vertical dashed green line in the middle of Fig.~\ref{fig2}),
where the relative gain increases from $\Gamma = -1.0$ (no gain)
to $\Gamma = 0$ (fully compensated cavity) and $\Gamma = 0.99$
(gain-dominated cavity). As discussed above, not only does the
number of observable Rabi oscillations increase, but they are
also decelerated (i.e., their period increases), while eventually
the populations exceed unity as expected.

What one can immediately observe in the dynamics of Fig.~\ref{fig5}(a)
is that the period of the oscillations follows an inverse-square dependence
on $\Gamma$, as expected from Eq.~(\ref{Eq:omegadif}), and as one can 
retrieve by calculating specific examples of dynamics. This suggests a way
to deduce the period of Rabi oscillations in ultrafast QE--cavity systems,
where the oscillations, with periods of a few fs, cannot be resolved with
current instruments. But in a loss-compensated cavity, where the gain is
externally controlled, one can increase the period of the oscillations to
the point where the instrument resolution allows clear observation of the
dynamics, and then extrapolate to the expected value in the absence of gain.
Gain-dominated cavities provide thus the means to not only manipulate the
dynamics of the QE--cavity system, but potentially also characterise it
through gradual modification of the provided gain.

\section{Conclusion}

We have analysed the dynamics of QEs coupled to optical
cavities that can be controlled via externally-provided
gain. Based on a general Rabi-visibility criterion that
quantifies the number of oscillations one should expect
to observe in an experiment, we established three different
coupling regimes, namely weak, strong, and coalescing. We
showed that the provided gain affects differently these three
regimes. When the coupling strength ($G$) is large,
the dynamics (and particularly the period of Rabi oscillations)
is not considerably affected by gain, and only the populations
of the ground and excited states change, being allowed to exceed
unity. On the other hand, for weak and intermediate coupling
strengths, externally pumping the system eventually governs the
dynamics, and the period of the Rabi oscillations increases,
suggesting that one could use gain to resolve ultrafast dynamics
by controllably and reversibly accelerating and decelerating the
oscillations. Finally, in the
coalescing regime, near the EP of the resulting $\pt$-symmetric
cavity, all system dynamics is completely governed by gain,
both populations oscillate nearly in phase, and they are allowed
to dramatically exceed unity. Such dynamics opens up new
possibilities for the design of dynamically controlled cavities
for strong-coupling realisations.

\section*{Acknowledgments}

We thank P. Edderkop for assistance with preparation of the graphs
and the web design.
C.~W. and F.~T. acknowledge funding from MULTIPLY fellowships under the
Marie Sk\l{}odowska-Curie COFUND Action (grant agreement
No. 713694).
M.~H.~E. acknowledges funding from Independent Research Fund Denmark (Grant No. 0165-00051B).
N.~A.~M. is a VILLUM Investigator supported by VILLUM Fonden
(grant No. 16498).
The Center for Polariton-driven Light--Matter Interactions (POLIMA) is sponsored by the Danish National Research Foundation (Project No.~DNRF165).

\hfill

\appendix

\section{Details on time evolution}\label{app:A}

The dynamics of the coupled QE--cavity system are governed by the
linear system typified by Eq.~(\ref{Eq:Hamiltonian}). Under the
conditions considered in the main text (i.e., $\omegacav = 
\omegax = 0$) and using the dimensionless quantities introduced
in the same, we have 
\begin{subequations}
\begin{equation}
\frac{\partial}{\partial \tau} 
\begin{pmatrix} a(\tau) \\  b(\tau) \end{pmatrix} 
= 
 \begin{pmatrix}
  \Gamma & -i G \\
  -i G & - 1
 \end{pmatrix} 
 \begin{pmatrix}  a(\tau) \\ b(\tau) \end{pmatrix}
, \label{eq:H_t_a}
\end{equation}
which we write compactly as 
\begin{equation}
 \dot{\mathbf{x}}(\tau) = \mathbf{A} \, \mathbf{x}(\tau) 
 . \label{eq:H_t_b}
\end{equation}\label{eq:H_t}%
\end{subequations}
The determination of the time evolution of the state vector
$\mathbf{x} (\tau)$ can be straightforwardly calculated once
in possession of the eigenvalues ($\lambda_{\pm}$) and the
eigenvectors ($\mathbf{v}_{\pm}$) of $\mathbf{A}$, which read 
\begin{subequations}
\begin{equation}
 \lambda_\pm = \frac{\Gamma - 1}{2} \pm \frac{1}{2} \sqrt{(\Gamma+1)^2 - 4G^2} = \ci \Omega_\pm
  \label{eq:lambdaA_pm}
\end{equation}
and
\begin{equation}
\mathbf{v}_\pm = \mathcal{N}_\pm \left( \ci \frac{\Gamma + 1 \pm \sqrt{(\Gamma+1)^2 - 4G^2 }}{2G} , 1 \right)^\mathrm{T}
  , \label{eq:vecsA_pm}
\end{equation} \label{eq:eigensysA_pm}%
\end{subequations}
respectively; here, $\mathcal{N}_\pm$ are normalization constants.

Equipped with the eigenvalues and eigenvectors of $\mathbf{A}$ [cf. Eqs.~(\ref{eq:H_t})--(\ref{eq:eigensysA_pm})], the time evolution of the system follows from 
\begin{equation}
 \mathbf{x}(\tau) \equiv \begin{pmatrix}  a(\tau) \\ b(\tau) \end{pmatrix} 
 = C_1 \, e^{\lambda_{-} \tau} \, \mathbf{v}_{-} + C_2 \,e^{\lambda_{+} \tau} \, \mathbf{v}_{+} ,
\end{equation}
that is,
\begin{subequations}
\begin{align}
 a(\tau) &= C_1 e^{\lambda_{-} \tau} \mathcal{N}_{-} \left( \ci \frac{\Gamma+1-\eta}{2G} \right) \nonumber\\
 &+ C_2 e^{\lambda_{+} \tau} \mathcal{N}_{+} \left( \ci \frac{\Gamma+1+\eta}{2G} \right)  , \\[0.5em]
 b(\tau) &= C_1 e^{\lambda_{-} \tau} \mathcal{N}_{-} + C_2 e^{\lambda_{+} \tau} \mathcal{N}_{+} , 
 \end{align}%
\label{eqA:a_b_tau_1}%
\end{subequations}
where we have introduced $\eta = \sqrt{(\Gamma+1)^2 - 4G^2}$ for shorthand notation. The constants $C_{1,2}$ are determined by the initial conditions; hereafter, we assume that the emitter--cavity system is initially in the state $\mathbf{x}_0 \equiv \mathbf{x}(\tau_0=0) = \left( a(0) , b(0)  \right)^\mathrm{T} = \left( 0 , 1 \right)^\mathrm{T}$, corresponding to an empty cavity and all the population is in the emitter. The choice of these initial conditions implies that
\begin{equation}
 C_1 = \frac{\Gamma + 1 + \eta}{2 \eta \mathcal{N}_{-}} 
 \, , \qquad \qquad
 C_2 = -\frac{\Gamma + 1 - \eta}{2 \eta \mathcal{N}_{+}} 
 . \label{eqA:C12}
\end{equation}
Then, substituting Eqs.~(\ref{eqA:C12}) into Eqs.~(\ref{eqA:a_b_tau_1}), yields
\begin{subequations}
\begin{align}
 a(\tau) &= \frac{\ci G}{\eta} \left[ e^{\lambda_{-} \tau} - e^{\lambda_{+} \tau} \right]\, , \\
 b(\tau) &= \frac{\Gamma+1}{2\eta} \left[ e^{\lambda_{-} \tau} - e^{\lambda_{+} \tau} \right] + \frac{1}{2} \left[ e^{\lambda_{-} \tau} + e^{\lambda_{+} \tau} \right] \, .
 \end{align}
\label{eqA:a_b_tau_lambda_eta}
\end{subequations}
Moreover, noting that $\lambda_\pm = \ci \Omega_\pm$ and $\eta = \ci \Omega_\mathrm{R}$ as well as $1-\Gamma = \Omega_\mathrm{R}/\qrabi$, Eqs.~(\ref{eqA:a_b_tau_lambda_eta}) can be written as
\begin{subequations}
\begin{align}
 a(\tau) &= -\frac{2\ci G}{\Omega_\mathrm{R}} \sin\left(\frac{\Omega_\mathrm{R}}{2}\tau  \right) e^{-\frac{\Omega_\mathrm{R} \tau}{2\qrabi}} \, , \\
 b(\tau) &= \left\{ %
 -\frac{\Gamma+1}{\Omega_\mathrm{R}} \sin\left(\frac{\Omega_\mathrm{R}}{2}\tau\right)
 + \cos\left(\frac{\Omega_\mathrm{R}}{2}\tau\right) 
 \right\} e^{-\frac{\Omega_\mathrm{R} \tau}{2\qrabi}}  \, ,
 \end{align}%
\label{eqA:a_b_tau_gamma_Rabi}%
\end{subequations}%
which are the sought-after amplitudes governing the time evolution of the coupled QE--cavity system [cf.~Eqs.~(\ref{Eq:Dynamics})].

\bibliography{references}

\end{document}